\documentclass[twocolumn,showpacs,preprintnumbers,amsmath,amssymb]{revtex4}
\usepackage{amsmath}
\usepackage{amsthm}

\usepackage{graphicx}
\usepackage{dcolumn}
\usepackage{bm}

\newtheorem{theorem}{Theorem}
\newtheorem{definition}{Definition}
\newtheorem{corollary}{Corollary}

\begin{document}

\title{Mutual information is copula entropy}

\author{Jian Ma}
\author{Zengqi Sun}%
 \email{majian03@mails.tsinghua.edu.cn}
\affiliation{%
Department of Computer Science, Tsinghua University, \\ Beijing 100084, China\\
}%

\date{\today}

\begin{abstract}
We prove that mutual information is actually negative copula
entropy, based on which a method for mutual information estimation
is proposed.
\end{abstract}

\maketitle

\section{Introduction}
In information theory, mutual information (MI) is a difference
concept with entropy.\cite{cover1991eit} In this paper, we prove
with copula \cite{nelsen1999intro} that they are essentially same --
mutual information is also a kind of entropy, called \emph{copula
entropy}. Based on this insightful result, We propose a simple
method for estimating mutual information.

Copula is a theory on dependence and measurement of
association.\cite{nelsen1999intro} Sklar \cite{sklar59} proved that
joint distribution $D$ can be represented with copula $C$ and
margins $F$ in the following form:
\begin{equation*}
D(\mathbf{x})=C(F_1(x_1),\ldots,F_N(x_N)).
\end{equation*}
Derived by separating the margins from joint distribution, copula
has all the dependence information of random variables, which is
believed that mutual information does as well.

Here gives notation. $C,c$ denote copula function and copula
density; $D,F$ denotes joint distribution and marginal distribution;
$H,I,H_c$ denote entropy, mutual information, and copula entropy
respectively. Finally, bold letters represent vectors while normal
letters single variable.

\section{Theorem and Proof}
For clarity, we give directly the main results. Please refer to
\cite{nelsen1999intro} and references therein for more about copula.
\begin{definition}[Copula entropy]
Let $\mathbf{X}$ be random variables with marginal function
$\mathbf{u}$ and copula density $c(\mathbf{u})$. Copula entropy of
$\mathbf{X}$ is
\begin{equation}\label{ce}
    H_c(\mathbf{x}) = - \int_{\mathbf{u}}{c(\mathbf{u})\log{c(\mathbf{u})d\mathbf{u}}}.
\end{equation}

\end{definition}
\begin{theorem}\label{thm}
Mutual Information of random variables equals to the negative
entropy of their corresponding copula function:
\begin{equation}\label{eq:thm}
    I(\mathbf{x}) = - - H_c(\mathbf{x}).
\end{equation}
\end{theorem}
\begin{proof}

\begin{equation*} \label{eq:proof}
\begin{split}
I(\mathbf{x})&= \int_{\mathbf{x}}{p(\mathbf{x})\log{\frac{p(\mathbf{x})}{\prod_{i}{p_i(x_i)}}}d{\mathbf{x}}} \\
 &= \int_{\mathbf{x}}{ c(\mathbf{u}_x)\prod_{i}{p_i(x_i)} \log{c(\mathbf{u}_x)} d{\mathbf{x}}}\\
 &= \int_{\mathbf{u}_x}{ c(\mathbf{u}_x) \log{c(\mathbf{u}_x)} d{\mathbf{u}_x} }\\
 &= - H_c(\mathbf{x})
 \end{split}
 \end{equation*}

\end{proof}

Entropy is the information contained in joint density function and
marginal densities, while copula entropy is the information
contained in copula density. The following corollary show the
relation between them.
\begin{corollary}\label{cor}
\begin{equation}\label{eq:cor}
H(\mathbf{x}) = \sum_{i}{H(x_i)} + H_c(\mathbf{x})
\end{equation}
\end{corollary}
\begin{proof}
The corollary is an instant result from the definition of mutual
information and theorem \ref{thm}.
\end{proof}

The worthy-a-thousand-words results cast insight into the inner
relation between mutual information and copula, and hence builds a
connection between information theory and copula theory.

\section{Estimating Mutual Information via Copula}
With theorem \ref{thm}, we propose the methods of estimating mutual
information from data. The estimation composes of two steps:
\begin{enumerate}
\item estimating empirical copula density;
\item estimating copula entropy.
\end{enumerate}

Given samples $\{\mathbf{x}_1,\ldots,\mathbf{x}_T\}$ i.i.d.
generated from $X=[x_1,\ldots,x_N]$, we can easily derive empirical
copula density using empirical functions
\begin{equation}\label{eq:emp}
F_i(x_i) = \frac{1}{T}\sum_{i=1}^{T}{\chi(X_t^i \leq x_t^i)}
\end{equation}
where $i=1,\ldots,N$, and $\chi$ is indicator function. Let
$\mathbf{u}=[F_1,\ldots,F_N]$, and then we can derive a group of
samples $\{\mathbf{u}_1,\ldots,\mathbf{u}_T\}$ from empirical copula
density $\hat{c}(\mathbf{u})$.

Since entropy estimation is a much contributed topic, copula entropy
can be achieved by well-established methods. In the following
experiment, we adopt the method proposed by \cite{kraskov2004emi}.

\section{Experiments}
\begin{figure}\label{fig1}
  \includegraphics[width= 80mm]{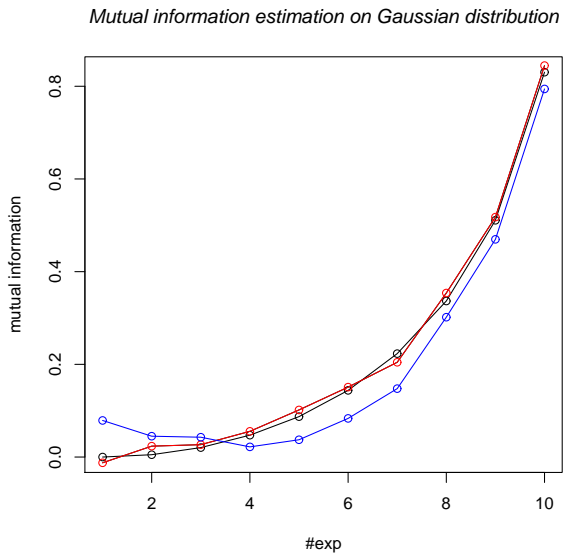}\\
  \caption{Mutual information estimation. Black represents analytic value,
  red represents by knn, and blue represents by copula entropy.}
\end{figure}

To evaluate the new estimation method, we consider tow correlated
standard Gaussian variables with covariance $\rho$, of which mutual
information is $-\frac{1}{2}\log{(1-\rho^2)}$. In the experiments
$\rho$ is set from 0 to 0.9 with step 0.1. With each value, we
generate a 1000 samples set. Copula entropy is estimated by the
method in \cite{kraskov2004emi}. As a contrast, mutual information
estimation in \cite{kraskov2004emi} was also run on the same sample
set. Figure \ref{fig1} illustrates the experimental results. It can
be learned that our method provides a competitive way of MI
estimation.

\newpage 
\bibliography{apssamp}

\end{document}